# *Photon Absorption Remote Sensing Virtual Histopathology: A Preliminary Exploration of Diagnostic Equivalence to Gold-Standard H&E Staining in Skin Cancer Excisional Biopsies*

Benjamin R. Ecclestone [1], James E. D. Tweel [1], Marie Abi Daoud [2], Hager Gaouda [1], Deepak Dinakaran [3], Michael P. Wallace [1], Ally Khan Somani [4,5], Gilbert Bigras [6], John R. Mackey [6,7], and Parsin Haji Reza [1]

**[1] University of Waterloo, 200 University Ave W, Waterloo, ON N2L 3G1, Canada**
**[2] University of Calgary, 2500 University Dr NW, Calgary, AB T2N 1N4, Canada**
**[3] University of Toronto, 27 King's College Cir, Toronto, ON M5S 1A1, Canada**
**[4] Indiana University School of Medicine, 550 N. University Blvd., UH 3240 Indianapolis, IN 46202**
**[5] SkinMD LLC, 16105 S. LaGrange Road, Orland Park. IL 60467**
**[6] University of Alberta, 116 St & 85 Ave, Edmonton, AB T6G 2R3, Canada**
**[7] illumiSonics Inc, 22 King St S Suite #300, Waterloo, ON N2J 1N8, Canada**

## Abstract

**Photon Absorption Remote Sensing (PARS) enables label-free imaging of subcellular morphology by observing biomolecule specific absorption interactions. Coupled with deep-learning, PARS produces label-free virtual Hematoxylin and Eosin (H&E) stained images in unprocessed tissues. This study evaluates the diagnostic performance of these PARS-derived virtual H&E images in benign and malignant excisional skin biopsies, including Squamous (SCC), Basal (BCC) Cell Carcinoma, and normal skin. Sixteen unstained formalin-fixed paraffin-embedded skin excisions were PARS imaged, virtually H&E stained, then chemically stained and imaged at 40x. Seven fellowship trained dermatopathologists assessed all 32 images in a masked randomized fashion. Example PARS and chemical H&E whole-slide images used in this study are publicly available at the BioImage Archive (https://doi.org/10.6019/S-BIAD2324).**

**Concordance analysis indicates 95.5% agreement between primary diagnoses rendered on PARS versus H&E images (Cohen's k=0.93). Inter-rater reliability was near-perfect for both image types (Fleiss' k=0.89 for PARS, k=0.80 for H&E). For subtype classification, agreement was near-perfect 91% (k=0.73) for SCC and was perfect for BCC. When assessing malignancy confinement (e.g., cancer margins), agreement was 92% between PARS and H&E (k=0.718). During assessment dermatopathologists could not reliably distinguish image origin (PARS vs. H&E), and diagnostic confidence was equivalent between the modalities.**

**Inter-rater reliability for PARS virtual H&E was consistent with reported benchmarks for histologic evaluation. These results indicate that PARS virtual histology may be diagnostically equivalent to traditional H&E staining in dermatopathology diagnostics, while enabling assessment directly from unlabeled, or unprocessed slides. In turn, the label-free PARS virtual H&E imaging workflow may preserve tissue for downstream analysis while producing data well-suited for AI integration potentially accelerating and enhancing the accuracy of skin cancer diagnostics.**

# 1. Introduction

Skin biopsies are a common surgical procedure performed when a suspicious skin lesion is identified (e.g., lesions with high-risk features of irregular borders, rapid growth or ulceration/bleeding), or to confirm a benign diagnosis for a lesion that requires resection due to local irritation or functional impairment. During a biopsy, part or all of the lesion is surgically removed.[1,2] Excised tissues are then assessed through histopathological methods to confirm or rule out malignancy.[1–4] For cancerous lesions, a margin assessment is also performed to identify the adequacy of resection.[3–5] Tissue margins are typically characterized as negative margins (edge of sample uninvolved with malignancy) or positive margin where cancer is suspected to be left behind on the patient.[3–5] Both the histology and margins are important for guiding further management. For example, basal cell carcinomas with positive margin could be safely monitored, whereas deeply invasive melanomas may require further oncologic resection locally as well as imaging and sentinel lymph node assessment.[3–5] Subsequently, the skin biopsy process may typically result in a definitive diagnosis, while achieving negative margins (complete removal) for smaller lesions, serving both diagnostic and therapeutic purposes.[3–5]

To enable histological analysis, skin excision samples undergo a standard tissue processing and staining procedure. Specimens are fixed in formalin, dehydrated, and embedded into paraffin wax. Embedded tissues are cut into thin translucent sections, placed onto slides, and stained with hematoxylin and eosin dyes (H&E),[6] the gold standard stain for histological evaluation. Hematoxylin stains cell nuclei in blue-purple, while eosin colors extracellular and cytoplasmic components in shades of pink-red.[6] This cellular microanatomy is critical for distinguishing malignant and benign skin conditions, as well as to determining the type and grade of cancer where applicable.[3–5,7]

Despite the diagnostic value, conventional tissue processing and labelling impart certain burdens. Chemical processing and staining are time intensive, carry significant costs, and require technical expertise.[6,8,9] Chemical labelling also imparts critical diagnostic limitations since current staining processes alter specimens' chemistry. As a direct result, sequential staining or "re-staining" of slides is normally prohibited in clinical diagnostic settings. Producing secondary staining contrasts beyond H&E (e.g., Toluidine Blue, Periodic Acid Schiff, Jones Silver) or repeating an H&E stain which fails to meet diagnostic standards (due to inter-sample variability factors such as section thickness, staining protocols, and reagent quality)[6,8,9], requires a new tissue section. Repeated sectioning directly constrains the diagnostic utility of specimens, especially when there are small tissue volumes (e.g., core biopsy samples). In the worst case, there may be a need to perform a re-biopsy to collect additional samples for staining.

As an alternative, there is growing interest in label-free nondestructive imaging technologies which can replicate histologic detail through fully digital "virtual staining" of unprocessed specimens. Virtual staining uses deep learning-based image-to-image transforms to emulate chemical staining contrast from a label-free image.

One promising label-free microscopy and virtual staining technology is Photon Absorption Remote Sensing (PARS).[10–13] PARS is an emerging imaging platform which observes the dominant radiative (fluorescent) and non-radiative (thermal and pressure) relaxation signatures following an optical absorption event. This method uses a pump-probe design, where the pump excites specimens using a focused picosecond-scale pulse. [10–13] Then, radiative relaxation is measured as the emission of Stokes shifted photons, while non-radiative relaxation is measured as transient fluctuations in the reflected or transmitted intensity of the co-focused probe beam. By simultaneously capturing both dominant de-excitation fractions, PARS provides unique insights into biomolecules' excited state dynamics and directly detects key biological molecules. [10–13]

In label-free histological imaging PARS employs (266 nm) ultraviolet (UV) excitation, to provide sensitivity to numerous diagnostic biomolecules such as hemoglobin, DNA, collagen, elastin, cytochromes. [10–15] The resulting PARS absorption data can subsequently identify a range of biomolecular features analogous to chemical staining. This forms a robust basis for deep learning-based PARS virtual staining. In practice, registered and paired PARS and chemically stained (e.g., H&E) whole-slide image datasets are used train image-to-image translation networks, such as Pix2Pix.[16] This process has been used to develop

PARS-based virtual staining for a range of chemical stains including H&E, PAS, Masson's Trichrome, and Jones Silver. [11,13,14,17,18] Prior work has demonstrated that these PARS virtually stained images may be effectively indistinguishable from chemical staining within the same tissue samples. [11–14,17] For instance, one recent study of PARS virtual H&E staining of breast core samples found significant pathologist agreement on diagnostic interpretation.[12]

If translated into clinical workflows, PARS virtual histology subsequently offers capabilities which may enhance current diagnostic processes. PARS avoids key chemical staining limitations by capturing endogenous biomolecular contrast label-free. This means that PARS imaging does not chemically or mechanically alter specimens, unlike chemical staining. Following PARS scanning each tissue specimen is completely preserved and can be reused for additional testing (e.g., immunohistochemical or secondary histochemical analysis). As a further advantage, the rich biomolecular specific data captured in PARS typically holds more information than afforded by traditional chemical H&E staining. Hence, one PARS scan, completed within a similar time to brightfield slide scanning, can replicate several additional histochemical stains beyond H&E (e.g., PAS, Masson's Trichrome, Jones Silver). [14,17,19] Subsequently, PARS virtual histology may greatly enhance the diagnostic utility of each tissue specimen by enabling (1) multiplexed virtual staining, and (2) sequential chemical staining of each tissue sample after PARS imaging. This may simultaneously advance diagnostic timelines by circumventing chemical staining processes. Finally, as endogenous biomolecules are leveraged for contrast, PARS (and other similar techniques) has shown preliminary label-free imaging within bulk resected tissue specimens. [20,21] This has potential to circumvent the entire sample preparation workflow in the future.

In this prospective study, we focus on evaluating the diagnostic accuracy of the PARS virtual H&E imaged against gold standard chemical H&E staining. This is a key validation step towards the ultimate goal of introducing PARS into clinical settings. This study specifically assesses diagnostic concordance between PARS virtual H&E staining and chemical H&E-staining, in whole slide excisional skin biopsy samples. Sixteen whole slide formalin-fixed paraffin-embedded (FFPE) skin excision samples were analyzed, representing normal skin, basal cell carcinoma (BCC), and squamous cell carcinoma (SCC). Unstained specimens were scanned using PARS to produce virtual H&E images. The same slides were then chemically H&E stained and imaged using a 40x digital pathology scanner, producing a total set of 32 whole slide images (16 PARS virtual H&E and 16 chemical H&E). For each primary diagnostic included in this study (SCC, BCC, benign), a matched PARS virtual and chemical H&E example are publicly available at the BioImage Archive [22] (see Data Availability). Seven fellowship-trained dermatopathologists independently reviewed these whole-slide images in a masked fashion and filled in diagnostic surveys. Concordance analysis is performed to quantify the diagnostic agreement, malignancy confinement, and diagnostic confidence between the PARS virtual H&E images and the gold standard H&E-stained samples.

# 2. Materials and Methods

### *2.1. Patient Materials*

A total of twenty-one independent cases of skin excisions were initially screened. A subset of sixteen formalin-fixed paraffin-embedded (FFPE) samples were selected to ensure representation across major histologic types. Five samples exhibited BCC, eight had SCC, and three were normal or benign skin tissue. Cases outside the study's scope (e.g., fungal infection) or redundant diagnoses were excluded.

The skin tissues were procured through collaboration with clinical partners. Samples were chosen from archival tissues no longer necessary for patient diagnostics, with all patient identifiers and all sample information removed. All human tissue experiments were conducted in accordance with the government of Canada TCPS2 guidelines. Specific study protocols were approved by the University of Waterloo Health Research Ethics Committee (Protocol ID: 40275; Photon Absorption Remote Sensing Microscopy of Surgical Resection, Needle Biopsy, and Pathology Specimens). As samples were fully anonymized archival tissues, the ethics committees waived informed consent.

### *2.2. Sample Preparation Prior to PARS Imaging and Gold Standard H&E Staining*

Skin lesions were surgically excised from patients and immediately fixed in 10% neutral buffered formalin for 24 hours. After fixation, a histotechnician prepared thin sectioned histopathology slides using a standard tissue processing and embedding workflow.[6] Specimens were dehydrated through graded ethanol rinses, cleared with xylene baths to remove ethanol and residual fats, and embedded into paraffin wax to form formalin-fixed paraffin-embedded (FFPE) tissue blocks. The FFPE blocks were sectioned with a microtome, into $\sim 4 - 5\ \mu\mathrm{m}$ translucent tissue sections, which were mounted on glass microscope slides and briefly heated to 60 °C to ensure adhesion and surface uniformity.

### *2.3. PARS Microscope Imaging*

Label-free PARS images were acquired from the unstained tissue sections using a custom-built transmission mode PARS microscope system. Detailed descriptions of the optical system, and imaging protocol are provided in a recent report by Tweel *et al*.[13] In brief, PARS imaging was performed using a 50 kHz 400 ps 266 nm UV laser (Wedge XF 266, RPMC; Bright Solutions, Pavia, Italy) to excite biomolecules. Following each excitation, three main features were measured representing the major absorption and scattering properties: radiative (fluorescence) emission, non-radiative heating (thermal/pressure effects), and local scattering.

To quantify the radiative signal intensity, emitted Stokes shifted photons were spectrally isolated and measured with an avalanche photodiode (APD130A2; Thorlabs, Newton, NJ, USA). The radiative intensity value was recorded as the peak emission amplitude. To measure the scattering and non-radiative relaxation effects, a 405 nm continuous-wave probe beam was co-aligned with the excitation (OBIS-LS405; Coherent, Santa Clara, CA, USA). Non-radiative (photothermal, and photoacoustic) signals were measured by observing heat and pressure induced modulations in the transmitted probe beam intensity following excitation. The time-resolved non-radiative modulation was measured by integrating the modulation to determine a percentage change in probe transmission. Scattering signals were recorded by measuring the 405 nm probe transmission intensity prior to excitation (APD130A2; Thorlabs, Newton, NJ, USA).

The three optical contrasts from each excitation event then correspond to one co-registered pixel in the final image. To form images, three-axis mechanical stages moved the sample across the objective lenses in a raster ("s"-shaped) scanning pattern. Pixels or excitation events were spaced 250 nm apart achieving 40× digital imaging magnification. Matching optical resolution was provided using a 0.42 numerical aperture (NA) UV objective lens (NPAL-50-UV-YSTF; OptoSigma, Santa Ana, CA, USA) to focus both excitation and detection beams, while optical signals (probe light and radiative photons) were collected using a 0.7 NA objective lens (278-806-3; Mitutoyo, Aurora, IL, USA).

Whole slide images were formed by discretizing the ROI into 500 × 500 µm sections which were independently scanned using the highlighted method. At each section, an autofocusing operation was performed prior to scanning. Section data was then contrast matched and stitched together forming a single whole-slide image, using the pipeline outlined in by Tweel *et al*.[13] The result was a complete whole-slide image based on native tissue absorption characteristics without the need for exogenous labels.

### *2.4. Gold Standard H&E Staining and Digital Image Acquisition*

Following PARS imaging, all tissues samples underwent standard chemical H&E staining using standard histopathology protocols. Stained slides were digitized at 40× magnification using a transmission brightfield slide scanner (Morpholens 1; Morphle Digital Pathology, New York, NY, USA). This produced high-resolution images of the chemically stained tissue to serve as ground truth for both training and validation.

The availability of precisely aligned virtual and chemical H&E images enabled robust image-to-image model development and direct head-to-head diagnostic comparison. The use of the same slide for both modalities—PARS prior to staining, and H&E afterward—ensured matched cellular and architectural features, minimizing confounding due to sectioning artifacts or tissue heterogeneity. This method of matched image generation has been validated in previous studies.[10–13,23]

### *2.5. PARS Virtual H&E Colourization*

A conditional generative adversarial network, Pix2Pix,[16] was trained to convert PARS label-free data into virtual H&E images. This approach differs from the previous PARS virtual staining validation study, which used a cycle-consistent generative adversarial network (CycleGAN).[12,24] CycleGAN does not require registration between the ground truth, and target domain.[12,25] As a result, CycleGAN may provide more optimal virtual staining performance (e.g., superior image clarity and sharpness) in samples where it is difficult to provide pixel-level ground truth (such as the previously explored need-core biopsy samples).[12,16,24,25] However, if pixel-level registration is possible, Pix2Pix can provide more robust correlations between structure and color, facilitating an improved domain transformation and better virtual staining image quality.[16,24] The key challenge to date has been developing a sufficient volume of pixel-level registered whole slide image data for model training. To this end, this work utilizes a new WSI registration workflow, based on the Warpy framework.[23] This enables pixel-level whole slide registration, facilitating successful implementation of the Pix2Pix virtual staining models.

In this study, two Pix2Pix virtual staining models were trained and applied to colorize the 16 whole slide virtual H&E images. Model 1 excluded Group A slides during training and was applied only to Group A; Model 2 excluded Group B and was applied to Group B (Figure 1a). This ensured that colorization was performed only on unseen data, supporting robust validation.

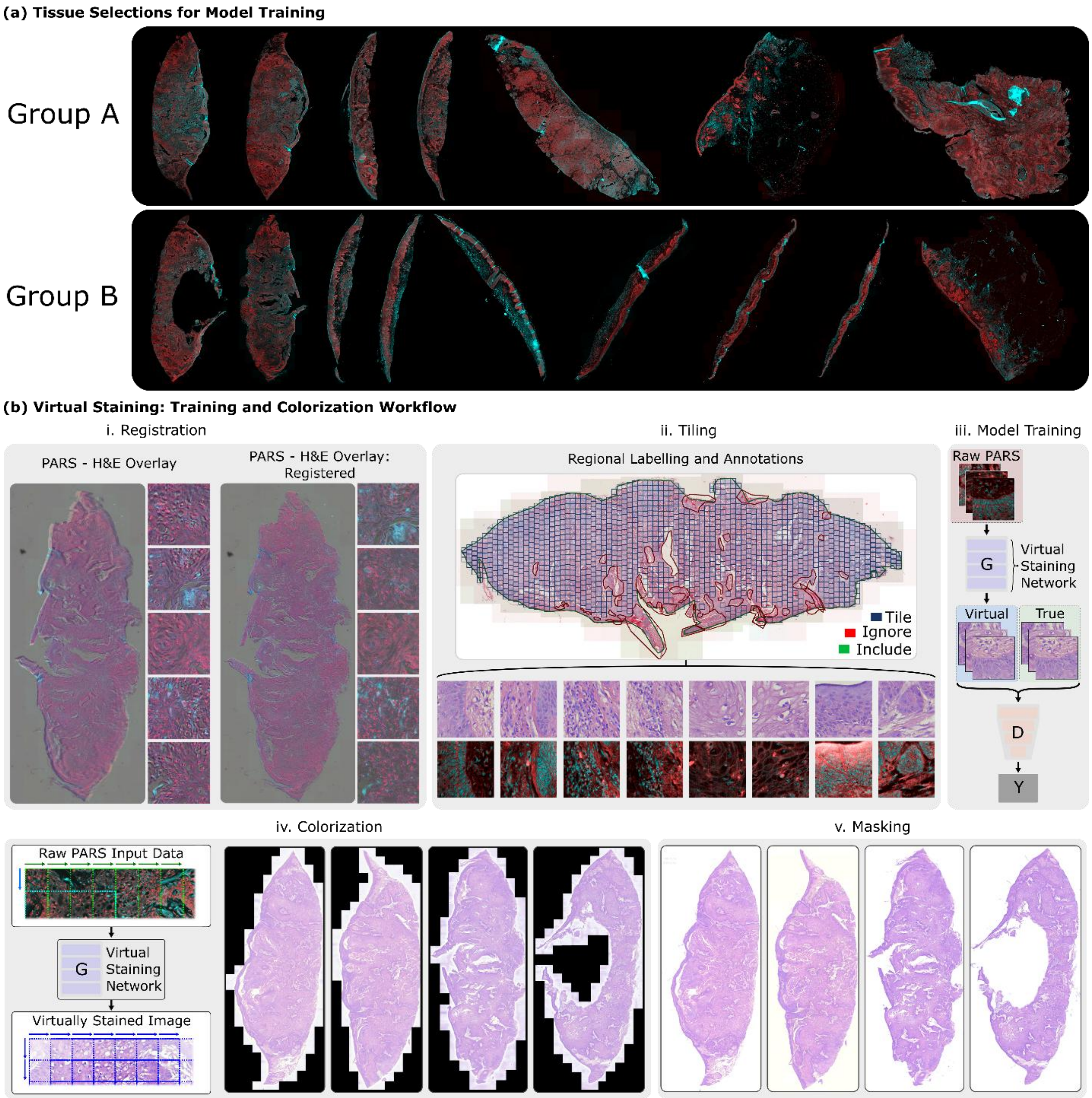

***Figure 1.*** *Workflow for developing PARS virtual H&E images. (a) Division of data for network training and colorization. Model (1) excluded data from Group A for training, while Model (2) excluded data from Group B during training. Models were then applied to virtually stain the respective groups ensuring only unseen data was colorized. (b) Steps performed to train and colorize tissues including, (i) whole slide registration using the Warpy whole slide registration workflow,[26] (ii) data labelling and (512 x 512 pixel) training pair generation, (iii) algorithm training, (iv) colorization or model application using 75% linearly blended overlapping regions, and (v) masking to conceal unscanned areas.*

Preparing the two Pix2Pix virtual staining models and developing the virtually stained images involved the following key steps (Figure 1(b)) (i) registration, (ii) tiling, (iii) training, (iv) colorization, (v) masking. Whole slide images (H&E and PARS) were registered using the Warpy workflow, which features manual rigid registration, automated affine registration, and a patchwise (500 × 500 µm patches) spline registration.[26] An example of slide overlays before and after registration are shown in Figure 1(b-i). Registered whole slide images were annotated to identify the tissue, and exclude artifacts (e.g., out of focus regions). Annotations were used to dice images into 512 × 512-pixel (128 × 128 µm) tile pairs (Figure 1(b-ii)) for model training. Model (1) used a total of 11459 training pairs, while Model (2) used a total of 11232 training pairs. Training was performed over 200 epochs, on a Nvidia RTX A6000 using pytorch 2.3.1 and Cuda 11.8 (Figure 1(b-iii)). Model (1) was applied to virtually stain the 7 unseen slides of Group (A), while Model (2) was applied to the 9 unseen slides from Group (B), resulting in 16 colorized whole slide images (Figure 1(b-iv)). During the colorization the model was applied to overlapping 512 × 512-pixel patches, which were linearly blended. Both PARS virtual H&E, and the ground truth H&E images were then masked (Figure 1(b-v)) to hide unscanned regions and remove background features (e.g., dust) which could indicate the image origin.

### *2.6. Evaluation by Expert Pathologists*

Seven fellowship trained dermatopathologists, six with no prior exposure to PARS virtual histology images, independently reviewed the 32 images (16 PARS and 16 H&E) using a customized web-based histopathology platform. Reviewers could freely navigate and zoom (up to 40×) each slide using the web interface. Pathologist scored each image on the parameters shown in Table 1.

Images were anonymized, randomly rotated, and assigned alphanumeric reference labels of AA through BF for survey purposes. All clinicopathologic details of the cases and the origin of the digital images (either true chemical H&E or PARS virtual H&E) were masked. Slides were presented in the following algorithmically randomized order: T16, T13, P8, T2, P4, P9, T14, P3, T6, T15, T12, P5, T11, P7, T1, T3, T10, P14, T8, T4, P11, T9, P6, P12, T7, P10, P15, P16, P13, P1, T5, P2. The 'P' prefix corresponds to PARS virtual H&E, while the 'T' prefix corresponds to true chemical H&E. This presentation sequence was designed using an algorithm which randomized image order while ensuring independent PARS and true H&E images from the same tissue were presented at least 7 images apart. The aim was to minimize potential for biases within each image pair, which might arise if (for example) an image pair from the same tissue were shown back-to-back.

***Table 1.*** *Survey questionnaire given to pathologists for each of the 32 images.*

| | | |
|---|---|---|
| ***1.*** | The primary tissue diagnosis is: | ☐ Non-Malignant/ Benign<br>☐ Squamous Cell Carcinoma<br>☐ Basal Cell Carcinoma<br>☐ Melanoma |
| ***2.*** | Squamous Cell Carcinoma | ☐ in situ SCC only<br>☐ well differentiated SCC<br>☐ moderately differentiated SCC<br>☐ poorly differentiated SCC |
| ***3.*** | Basal Cell Carcinoma | ☐ Low Risk * <br>☐ High Risk † |
| ***4.*** | Melanoma | ☐ High Dermal Mitotic Rate<br>☐ Skin Ulceration<br>☐ Desmoplastic melanoma<br>☐ Lymphatic invasion<br>☐ Brisk Tumor Infiltrating Lymphocytes |
| ***5.*** | Malignancy confined within the boundaries of the tissue in this section / margins negative | ☐ Yes<br>☐ No |
| ***6.*** | This digital image comes from H&E | ☐ Yes<br>☐ No<br>☐ Uncertain |
| ***7.*** | My diagnostic confidence is | ☐ Low<br>☐ Moderate<br>☐ High |

* e.g., nodular, superficial, pigmented, infundibulocystic, fibroepithelial
† e.g., basosquamous, sclerosing/morpheaform, keloidal, infiltrative, sarcomatoid

### *2.7. Statistical Analysis*

Agreement between PARS and H&E diagnoses was quantified using Cohen's kappa ($\kappa$) (for intra-rater modality comparison) and Fleiss' kappa ($\kappa$) (for inter-rater agreement within each modality). The Cohen's kappa coefficient measures inter-rater reliability while accounting for the possibility of agreement occurring by chance.[27] This is key to assessing if two raters agree beyond what would be expected by chance alone. The Kappa assessment also provides robustness to varied prevalences in different data categories.[27] The Fleiss' kappa measures the same characteristic as the Cohen's kappa but compares more than two raters.[28]

Kappa values range from −1 to 1. Positive values indicate agreement beyond chance (1 indicates perfect agreement), while negative values indicate agreement worse than chance. Kappa values were interpreted as "fair" or "poor" for $\kappa < 0.6$, "substantial" or "good" for $\kappa = 0.6 - 0.8$, "near perfect" for $\kappa > 0.8$.[29,30] Percent agreement metrics were also reported to contextualize kappa scores. All calculations were performed in R statistical software (version 4.2.0).[31]

## 3. Results

### *3.1. Example Whole-Slide Image Pairs*

Representative whole-slide images from two skin excision specimens are shown in Figures 2 and 3. For each tissue sample, the raw total absorption (TA) PARS image is presented alongside the PARS virtual H&E, and real H&E image. In the TA PARS image, the radiative relaxation is colored in blue, while the non-radiative relaxation is colored red. This image represents the raw PARS data which is inputted into the virtual staining algorithm. The virtual and chemical H&E images represent subsections of the diagnostic views shown to pathologists during assessment. Unfortunately, it is challenging to illustrate the virtual staining performance through these subsection examples due to the spatial limitations. Hence, a subset of the of whole-slide PARS and chemical H&E images have been made publicly available at the BioImage Archive (see Data Availability). Slides are a standard OME-TIFF format and can be viewed and compared with standard pathology viewing software such as QuPath.[32]

Within the presented examples, Figure 2(a) illustrates a case of basal cell carcinoma (BCC). This sample exhibits basaloid nests arising from the epidermis and within the dermis. High-magnification views (Fig. 2(a)i–ii) reveal classic basaloid nests and cords with hyperchromatic nuclei, scant cytoplasm, and peripheral nuclear palisading. These tumor clusters are embedded in a mucinous or fibroblastic stroma, consistent with nodular or superficial BCC. The absence of keratinization distinguishes BCC from squamous cell carcinoma, where mitotic figures are common but rarely atypical.

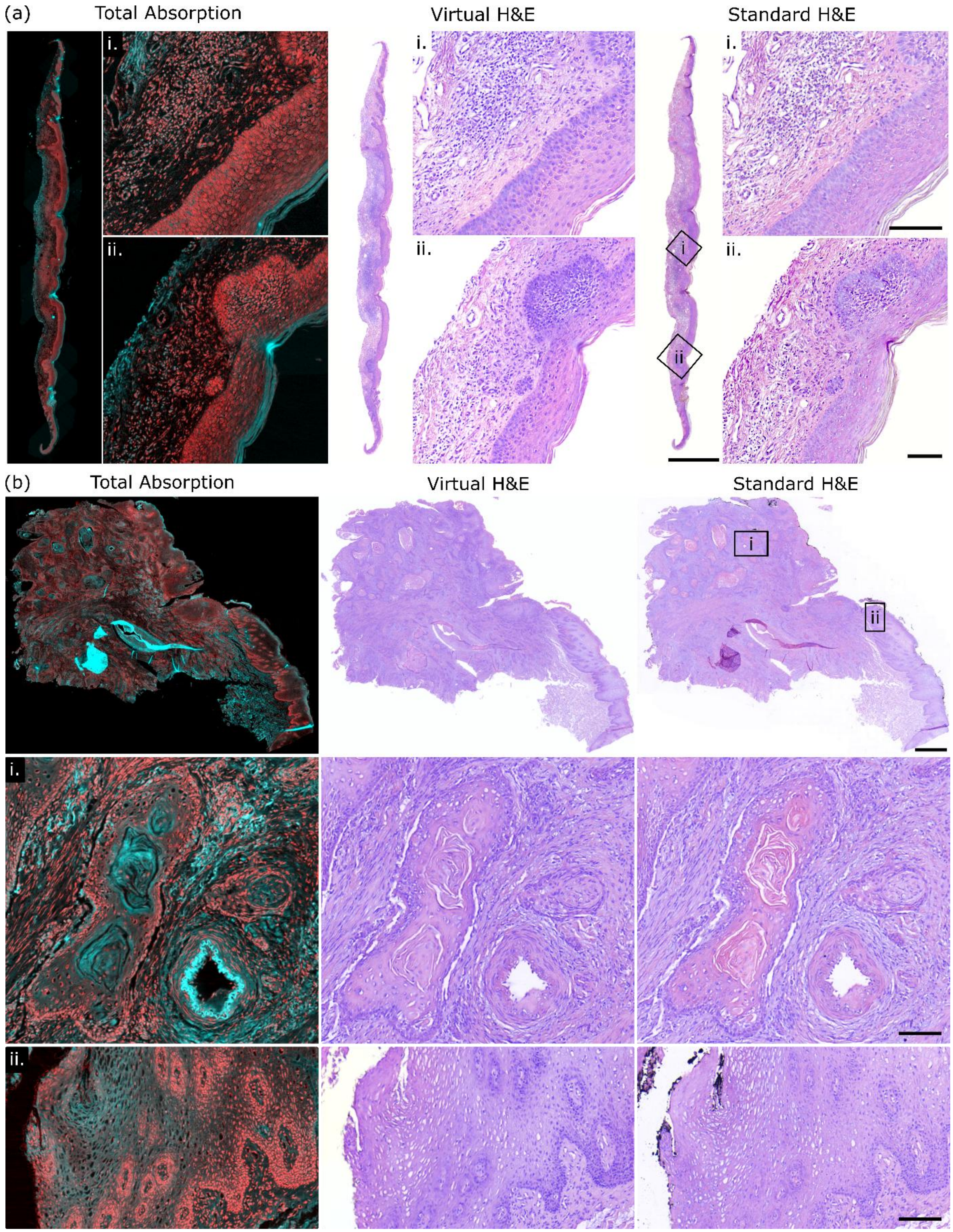

**Figure 2.** *Example PARS total absorption (TA), virtual H&E, and true H&E pair used in this study. The samples exhibit closely matched staining colours and tissue morphology between the virtual and real representations. (a) Example of basal cell carcinoma (BCC). (b) Example of squamous cell carcinoma (SCC). Sub figures (i,ii) depict two regions of higher magnification on the sample. (a) & b) Scale Bar: 1 mm, (i) & (ii) Scale Bar: 100 µm.*

Figure 2(b) depicts squamous cell carcinoma (SCC), arising from the epidermis and growing downward into the dermis. Higher magnification (Fig. 2(b)i–ii) shows atypical keratinocytes with pleomorphic, hyperchromatic nuclei and prominent keratinization. Features include dyskeratotic cells, keratin pearls, and intercellular bridges—all hallmarks of squamous differentiation. SCC invades the dermis in sheets, cords, or nests, with variable differentiation. Well-differentiated tumors show organized keratinization, while poorly differentiated lesions lack keratin and exhibit marked nuclear atypia and mitotic activity. Desmoplastic stroma may surround invasive nests, especially in aggressive variants.

Comparing the different visualizations, the "Total-Absorption" (TA) PARS images provide a clear basis for the success of the virtual staining. The deep UV excitation used in PARS is broadly absorbed in tissues, where the unique distribution of relaxation effects (radiative vs. non-radiative) provides specificity to critical diagnostic features. The non-radiative contrast (red) highlights predominately DNA which exhibits strong non-radiative relaxation, corresponding to similar features as hematoxylin staining. Conversely, the connective tissues (primarily collagen and elastin) exhibit appreciable radiative relaxation (blue) approximating eosin's affinity for cytoplasm and stroma.[10–13] In some instances, PARS may reveal additional features which are not visualized by the H&E staining. For example, the inner elastic wall of an artery exhibits prominent radiative signal (Figure 3 TA-i) not captured in standard H&E due to limited sensitivity to certain biomolecules. The virtual staining model suppresses these features to match the chemical H&E fidelity.

An additional advantage of the PARS virtual staining is contrast consistency. In PARS the signal (and virtual H&E stain contrast) is derived from the endogenous absorption. Hence, the virtual staining exhibits uniformity across all samples. In contrast, chemical H&E staining is subject to variability from factors such as section thickness, dye concentration, staining duration, specimen age, and storage conditions.[33–37] This improved consistency with virtual staining may support improved reproducibility in digital workflows.

As a further extension, while this work focuses only on virtual H&E staining, there are numerous other stains used in clinical settings. As an example, virtual staining networks could be trained to develop Periodic Acid Schiff and Toluidine Blue which are used to in certain cases of basaloid carcinomas.[38,39] In this work, H&E was selected to align with current practices for skin malignancy diagnosis. The turnaround for a pathology report is typically a week or more due to tissue preparation, chemical H&E staining, brightfield imaging, and human diagnostic interpretation. Hence, the additional stains beyond H&E are often avoided to mitigate further delays. However, each PARS scan of a single unstained slide generates rich, multidimensional molecular data that has the potential to replicate several chemical staining patterns. With the development of additional virtual staining models. Multiple stains may be developed from each tissue specimen. In addition, label-free imaging retains samples ancillary testing and analysis.

### *3.2. Primary Diagnosis*

Pathologists were first asked to assign a primary diagnosis for each image (PARS, and H&E), resulting in 32 (16 PARS and 16 H&E) preliminary diagnoses per rater, for a total of 224 primary diagnoses. Corresponding primary diagnoses for the PARS and H&E images are presented in Table 2. For reference, the expectation or expected diagnosis is determined from the most prevalent diagnosis for each tissue sample. Inter-rater agreement for the resulting primary diagnosis was high within each modality. There was 94.6% agreement (Fleiss' kappa $\kappa = 0.886$) within the PARS images, and 92.8% agreement (Fleiss' kappa $\kappa = 0.805$) within the H&E images. This indicates near perfect inter-rater agreement for diagnosis within the PARS and within the H&E images.

Of the total 224 primary diagnosis, this corresponds to a total of 14 disagreements of the primary diagnosis, 6 on the PARS images and 8 on the H&E images. Notably, 6 of the 14 disagreements in primary diagnosis occurred on a single sample. This corresponded to 3 disagreements in primary diagnosis (3 Benign vs. 4 BCC) on the PARS image, and 3 disagreements in primary diagnosis (3 Benign vs. 4 BCC) for H&E. Excluding this difficult to diagnose sample, the agreement in the primary diagnosis rises to 97.2% for the PARS images, and 95.4% for the H&E images.

***Table 2.*** *Summary of pathologist responses to question "The primary tissue diagnosis is".*

| | **PARS** | | **H&E** | | |
|---|---|---|---|---|---|
| **Diagnosis** | Response | Discrepancies | Response | Discrepancies | **Expectation** |
| SCC | 56 | 0 | 57 | +1 | 56 |
| BCC | 32 | -3 | 31 | -4 | 35 |
| Melanoma | 1 | +1 | 2 | +2 | 0 |
| Benign | 23 | +2 | 22 | +1 | 21 |
| **Total** | **106** | **(6)** | **104** | **(8)** | **112** |

Inter-modality agreement between the PARS and H&E was similarly strong. Across raters the average diagnostic concordance between PARS and H&E was 95.5%, with a Cohen's kappa ($\kappa = 0.93$) indicating near-perfect agreement. In total, basal' was chosen 63 times, 'non-malignant' 45 times, 'squamous' 113 times, and 'melanoma' 3 times. There were a total of 5 disagreements on diagnoses between an H&E and PARS pair. Three pathologists had perfect agreement ($\kappa = 1$), while the remaining four had near-perfect agreement ($\kappa = 0.8 - 0.9$). These results indicate reliable and equivalent diagnostic discrimination between BCC, SCC, and benign lesions across modalities for the specimens used in this study.

***Table 2.*** *Summary of pathologist responses to question "The primary tissue diagnosis is".*

| | | **H&E** | | | | |
|---|---|---|---|---|---|---|
| | Diagnosis | SCC | BCC | Melanoma | Benign | **Total** |
| **PARS** | SCC | 56 | - | - | - | **56** |
| | BCC | - | 30 | - | 2 | **32** |
| | Melanoma | - | - | 1 | - | **1** |
| | Benign | 1 | 1 | 1 | 20 | **23** |
| | Total | **57** | **31** | **2** | **22** | **112** |

### *3.3. Evaluation of Cancer Subtypes and Grading*

A similar analysis to the primary diagnosis response was performed to assess the concordance for the evaluation of the malignant subtypes, as per the Table 1 criterion for each of the cancers.

#### *3.3.1 Squamous Cell Carcinoma:*

SCC was diagnosed 113 times (the number is uneven due to a discordant diagnosis between the PARS and H&E as noted in the ***Primary Diagnosis*** section). The SCC subtyping responses were as follows 'well differentiated' (n = 68) or 'moderately differentiated' (n = 39), with 'poorly differentiated' (n = 3) and 'in situ' (n = 3) rarely chosen (Table 3).

Inter-rater agreement was measured using the 7 cases where there was complete agreement of the SCC primary diagnosis. The Fleiss' kappa for inter-rater agreement was computed as $\kappa = -0.0121$ for the H&E images, and $\kappa = 0.0392$ for the PARS images. This indicates no inter-rater agreement between pathologists within the H&E images, or within the PARS images.

Despite low inter-rater agreement, intra-rater agreement between matched H&E–PARS pairs showed 91% concordance with a mean Cohen's κ of 0.73. This suggests individual raters preserved consistent subtype interpretations across the PARS and H&E image pairs. As an aside, rater 1 exhibited a very low kappa ($\kappa = 0$) despite exhibiting a 5/7 success rate. This is an effect of calculating the Kappa statistic with a relatively low sample size.

***Table 3.*** *Summary of pathologist responses to the question of the Squamous Cell Carcinoma Subtyping.*

| Rater | Agreements | Total SCC Selections | Kappa |
|---|---|---|---|
| 1 | 5 | 7 | 0.0 |
| 2 | 6 | 8 | 0.5 |
| 3 | 9 | 9 | 1 |
| 4 | 8 | 8 | 1 |
| 5 | 8 | 8 | 1 |
| 6 | 7 | 8 | 0.619 |
| 7 | 8 | 8 | 1 |
| **Total** | **51** | **56** | **0.73** |

### *3.3.2 Basal Cell Carcinoma:*

BCC was selected as the primary diagnosis 63 times. Excluding 3 instances of discordant diagnosis between H&E and PARS image pairs (i.e., 'basal' was selected for one image and not the other (Table 2)), 60 matched image sets remained. Of these responses 58 gradings were 'low' and just 2 were 'high'. The two 'high' responses are attributed to Rater 4, on a matching set of PARS and H&E images. Therefore, the PARS and H&E yielded 100% agreement between matched virtual and chemical H&E for BCC grading.

### *3.3.3 Melanoma:*

Melanoma samples were not included within the dataset of this study. However, melanoma was added to the diagnostic survey to act as a distractor option. In total, melanoma was selected only three times (Rater 1: 2; Rater 6: 1), with subtype features reported once. This corresponds to a selection rate of 1.3% within the entire diagnostic set (of 224 preliminary diagnosis). Due to the low frequency, a Kappa value was not computed for these selections.

## *3.4. Evaluation of Cancer Margins/Tissue Edges*

If pathologists identified malignancy (BCC, SCC, Melanoma), they were asked to assess the resection margins. Malignancy confined within the boundaries of the tissue corresponds to negative margins indicating a complete resection. Of 179 malignant diagnosis, 178 responses were recorded: 140 “no” and 38 “yes”.

Inter-rater agreement was high at 90% within the H&E and PARS images, corresponding to a Fleiss’ kappa of $\kappa = 0.784$ for H&E images, and $\kappa = 0.824$ for PARS images. This was calculated using the 10 image pairs which exhibited universal malignancy diagnosis (i.e., all 7 raters selected BCC, or SCC). Other tissues were benign ($n = 3$) or had less than 100% agreement in the primary diagnosis ($n = 3$). Within this set of 10 tissues, all disagreements of margin assessment occurred on a single H&E and PARS pair. Responses were (4 ‘yes’: 3 ‘no’) for the H&E, and (6 ‘yes’: 1 ‘no’) for the PARS. This suggests that this sample was difficult to assess, potentially due to poor tissue selection or poor specimen quality. Excluding this case, there was perfect (100%) inter-rater agreement during margin assessment within the PARS and within the H&E images.

Inter-modality concordance was similarly high between the PARS and H&E during margin assessment. The average inter-modality agreement was 92%, with Cohen's $\kappa = 0.718$. This is calculated using the sample pairs where each pathologist had a concordant diagnosis. This indicates significant agreement of margin assessment between the PARS and H&E images.

***Table 4.*** *Summary of pathologist responses to the question of malignancy confinement.*

| Rater | Agreements | Total Margin Selections | Kappa |
|---|---|---|---|
| 1 | 10 | 10 | 1 |
| 2 | 11 | 13 | 0.409 |
| 3 | 12 | 14 | 0.44 |
| 4 | 13 | 13 | 1 |
| 5 | 12 | 12 | 1 |
| 6 | 10 | 12 | 0.429 |
| 7 | 11 | 12 | 0.75 |
| **Total** | **79** | **86** | **0.718** |

### *3.5. Image Origin*

Raters were asked to classify image origin as a chemical H&E, virtual (PARS) or uncertain (Table 5). The option “uncertain” was chosen most often (86/224), indicating pathologists were frequently unsure of the image origin. The fourth respondent selected ‘Uncertain’ for all 16 PARS and 16 H&E images. Among the other raters, there were 25 instances H&E images were misidentified as virtual (PARS), and 38 instances where virtual H&E was misidentified as real images.

Raters were only slightly more likely to consider H&E images genuine (40%) than virtual (34%). These suggest that pathologists could not reliably distinguish between conventional H&E and PARS virtual H&E.

***Table 5.*** *Summary of pathologist responses to question “This digital image comes from H&E”.*

| | | Rater | | | | | | | |
|---|---|---|---|---|---|---|---|---|---|
| **Image** | **Response** | 1 | 2 | 3 | 4 | 5 | 6 | 7 | **Total** |
| H&E | Genuine | 10 | 0 | 9 | 0 | 11 | 12 | 3 | **45** |
| | Uncertain | 1 | 11 | 0 | 16 | 2 | 0 | 12 | **42** |
| | Virtual | 5 | 5 | 7 | 0 | 3 | 4 | 1 | **25** |
| PARS | Genuine | 10 | 0 | 6 | 0 | 7 | 13 | 2 | **38** |
| | Uncertain | 3 | 9 | 2 | 16 | 2 | 0 | 12 | **44** |
| | Virtual | 3 | 7 | 8 | 0 | 7 | 3 | 2 | **30** |

### *3.6. Diagnostic Confidence*

Respondents were asked to rate their diagnostic confidence as Low, Moderate, or High (Table 6). Of 224 total responses, 167 were high (~74.5% of cases), 42 moderate (~18.75% of cases), and 15 low (~6.7% of cases). This corresponds to a diagnostic confidence of moderate or higher in 93% of the diagnostic cases. Inter-rater agreement for this diagnostic confidence was found to be fair within the PARS (Fleiss' kappa $\kappa = 0.235$), and within the H&E images (Fleiss' kappa $\kappa = 0.264$).

Inter-modality agreement was also calculated to compare raters diagnostic confidence between image pairs. The average Cohen's kappa across raters was $\kappa = 0.36$ indicating fair agreement of diagnostic confidence between the PARS and H&E images. There were 23 cases where a raters diagnostic confidence differed within a PARS and H&E image pair. Of the disagreements, 11/23 favored the PARS image, while 12/23 favored the H&E images.

***Table 6.*** *Summary of pathologist responses to the question "My diagnostic confidence is"*

| Score | H&E | PARS | Total |
|---|---|---|---|
| Low | 8 | 7 | **15** |
| Moderate | 20 | 22 | **42** |
| High | 84 | 83 | **167** |
| Kappa | 0.264 | 0.235 | |

# 4. Discussion

### *4.1. Study Summary and Key Findings*

This study evaluated the diagnostic performance of PARS-generated virtual H&E images compared to conventional chemical H&E staining in excised skin tissue specimens. Sixteen unstained FFPE skin excisional biopsies were imaged using PARS, virtually stained, then chemically labelled with H&E and digitally scanned at 40× magnification. The resulting 32 images were assessed by seven fellowship-trained dermatopathologists in a masked and randomized fashion. Pathologists evaluated freely evaluated diagnostic features across the entire whole slide images (with the image origin masked), at any magnification (up to 40x).

Across all diagnostic metrics (i.e., primary diagnosis, tumor subtype, and malignancy confinement) PARS images demonstrated significant concordance (>90% agreement, with a $\kappa > 0.7$) comparable to, or exceeding, chemical H&E evaluation of the same tissue section.

For the primary diagnosis, raters showed near perfect inter-rater agreement (94.6%) within PARS ($\kappa = 0.886$) and (92.8%) within H&E images ($\kappa = 0.805$). Inter-modality agreement was even stronger between the PARS and H&E images with a total 95.5% agreement (Kappa 0.93). Many of the disagreements were localized to a single sample, potentially indicating a difficult to diagnose specimen. As a further note, melanoma, which was not imaged in this case series but was given as a distractor option on the questionnaire, was only selected 3 times. This corresponded to a very low selection rate of ~1.3%. These findings indicate very strong concordance and confidence in the primary diagnosis. Substantial concordance was also achieved when malignancy subtype or grade. For SCC subtyping, there was 91% inter-modality agreement ($\kappa = 0.73$) between the PARS and H&E images. For BCC subtyping, there was perfect agreement amongst all raters. When assessing resection margins, there was substantial inter-rater agreement within the H&E ($\kappa = 0.784$), and PARS ($\kappa = 0.824$). There was also substantial inter-modality agreement (92% with $\kappa = 0.718$) between PARS and H&E.

Importantly, raters were unable to reliably distinguish virtual from real H&E images, and the diagnostic confidence was similar for each modality. In some measures, such as margin assessment (PARS: $\kappa = 0.824$, H&E $\kappa = 0.784$), or primary diagnosis (PARS: $\kappa = 0.886$, H&E κ=0.805), PARS even slightly outperformed the chemical H&E images. This suggests that PARS virtual staining may enable more consistency in diagnosis as compared to the H&E images. These differences may reflect an improvement in stain standardization, as virtual staining eliminates the variability associated with chemical protocols such as section thickness, reagent quality, and staining duration. [6,8,9] Overall, this analysis strongly supports the conclusion that PARS virtual H&E images provide diagnostic performance that matches or exceeds conventional H&E images.

### *4.2. Interpretation of Findings in Context*

While chemical H&E staining remains the diagnostic gold standard, it is subject to interobserver variability and process-related inconsistencies. Developing a diagnosis often requires assessment of subtle features of nuclear morphology and tissue organization.[40–43] This complicates the evaluation of new histologic techniques, such as PARS virtual H&E images, since the “gold standard” chemical H&E assessment is itself imperfect. In the context of this study, PARS virtual histology in skin excisions is shown to be diagnostically equivalent to chemical H&E staining assessed through digital histopathology.

Diagnostic concordance for common skin cancers such as BCC and SCC is generally high. [44,45] Well-defined and common diagnoses often report agreement exceeding 90% in many cases.[44,45] This aligns with our findings for the primary diagnosis, which exhibited high inter-rater agreement (PARS: $\kappa = 0.886$, H&E κ=0.805) and high diagnostic concordance (95.5% agreement, $\kappa = 0.93$). Though, performance can vary depending on the complexity or rarity of the lesion, and the pathologist specialization.[46]

In this study melanoma was added as a distractor option, on the survey and was only selected 3 times across the 224 primary diagnoses. This was selected twice for H&E images, and once on a PARS image. This corresponds to a 1.3% selection rate, further supporting the diagnostic confidence for the primary tissue diagnosis. As an additional confirmation step, post analysis was performed on the three examples where melanoma was selected to ascertain the potential rationale behind the selection. For all three examples, post analysis and pathologist consensus ruled out the presence of melanoma. However, sparse areas of keratinization and connective tissues were found to exhibit a more brown-yellowish tinge on certain monitors, as compared to other samples in this study. It is possible these features acted as a confounding factor during the primary diagnosis.

In this study, there was significant inter-modality agreement between the PARS and H&E (91%, $\kappa = 0.73$) for the SCC subtype, despite poor inter-rater agreement (H&E $\kappa = -0.0121$, PARS $\kappa = 0.0392$). Concurrently, the diagnostic confidence exhibited only fair inter-rater agreement (PARS: $\kappa = 0.235$, H&E: $\kappa = 0.264$), with a nearly identical distribution of confidence scores across PARS and H&E. This indicates that despite the potential variability between individual pathologists' interpretation PARS offers comparable interpretability and diagnostic confidence to chemical H&E images.

Despite high overall concordance for common skin cancers, variability remains a concern for rare subtypes or lesions at the interface of benign and malignant features. This illustrates the importance of new methods and improvements to diagnostic tools which may enhance reliability and accuracy. As an example, virtual staining could help to reduce pre-analytic variations. Unlike chemical staining, which is sensitive to section thickness, tissue processing, and reagent variability, PARS produces consistent stain outputs.[6,8,9] This standardization may explain the high inter-rater agreement observed for PARS across multiple diagnostic tasks. Coupled with emerging technologies in AI diagnostics, PARS has significant potential to enhance these diagnostic processes.[47–49]

***3. Strengths, Limitations, and Future Research***

Careful consideration was taken to minimize factors which could unduly influence pathologist responses. Pathologists were carefully masked to tissue origin. Images were randomly oriented and randomly ordered with an enforced minimum spacing between image pairs. Image backgrounds were removed to further mask the underlying modality. In addition, every effort was made to ensure that standard diagnostic practices were facilitated. Pathologists were allowed to freely assess the 40× digital histology images as they would during conventional workflows. Examples of these whole slide PARS and chemical H&E images are publicly available at the BioImage Archive (see Data Availability). Efforts were also taken to minimize confounding effects from sample variance. PARS and H&E images were derived from exactly same tissue sections, rather than adjacent slides, ensuring direct nuclei to nuclei pairing and identical diagnostic features in each image.

The conclusions presented here are encouraging but limited by sample size of tissues ($n = 16$ slides), and pathologists ($n = 7$ raters). The diagnostic scope also limits the generalization by focusing on the most common skin lesions (BCC and SCC).

Future studies should expand the number of raters and increase the number of samples to further generalize and expand upon the findings presented here. Other studies should also expand the diagnostic scope to incorporate melanocytic lesions and additional cancer subtypes. These works should also expand outside of skin malignancies to include other major cancer types, such as gastrointestinal cancers.

In addition, future studies should include additional histochemical stains. This is an exciting avenue of development, given PARS unique ability to produce multiplexed virtual histochemical visualizations from a single scan. [14,17,19]This offers a direct avenue to improve the diagnostic accuracy beyond chemical staining workflows, by allowing multiplexed virtual staining on each of the assessed specimens. Finally, future studies should focus on additional sample formats. While, the present study focuses on FFPE specimens, PARS has previously shown capability to image thick unstained specimens including bulk unprocessed tissues. [10,20] Future works may explore using PARS to provide virtual staining (e.g., H&E) directly within bulk unprocessed tissues (freshly resected, or preserved).

# 5. Conclusions

PARS label-free virtual histology imaging offers a promising alternative to chemical staining workflows. By observing the dominant radiative (fluorescent) and non-radiative (thermal) relaxation processes following an optical absorption event, PARS generates rich datasets which are a strong starting point for artificial intelligence imaging and diagnostic methods. This has enabled label-free PARS virtual staining, which is effectively indistinguishable from chemical staining, for a range of dyes (e.g., PAS, Masson's Trichrome, Jones Silver).[17,19]

This study finds that corresponding PARS based virtual H&E images are diagnostically equivalent to chemical H&E staining across major metrics. Pathologists interpreted PARS images with similar confidence and accuracy and could not reliably distinguish image origin. Hence, the results of this study add to a growing body of work which indicate that PARS virtual staining may be applicable to histological evaluation. These findings simultaneously serve as a critical clinical validation of the PARS virtual H&E contrast, moving towards the ultimate goal of generalized H&E diagnostic capacity.[11–13] In future, this PARS virtual staining may significantly enhance the diagnostic utility of specimens by reducing tissue consumption, providing multiplexed histology representation, and enabling ancillary diagnostic testing. This may eventually transform histological workflows by immediately reducing timelines and cost, while simultaneously producing datasets which enable seamless integration towards AI diagnostics.

Future extensions of this work include larger multi-institutional validation studies, exploration of broader staining repertoires, and exploration of additional malignancies and tissue types. Ultimately, as PARS directly targets endogenous contrast of the tissues, this technique can be applied to any form of tissue, including freshly resected specimens,[10] and living tissues in vivo.[50] As a result, the PARS virtual histology could be applied to image freshly resected tissues (e.g., Mohs surgical resections).

In aggregate, the high-fidelity virtual histology capabilities PARS imaging stands to transform the landscape of digital pathology.

***Institutional Review Board Statement:*** Tissues were acquired through collaboration with clinical partners. Specimens were acquired under protocols (Protocol ID: HREBA.CC-18-0277) with the Research Ethics Board of Alberta and (Photoacoustic Remote Sensing Microscopy of Surgical Resection, Needle Biopsy, and Pathology Specimens; Protocol ID: 40275) with the University of Waterloo Health Research Ethics Committee. All human tissue experiments were conducted in accordance with the government of Canada guidelines and regulations, including "Ethical Conduct for Research Involving Humans (TCPS2)".

***Informed Consent Statement:*** The samples were obtained from anonymous patient donors, with all patient identifiers removed to ensure anonymity. The ethics committee waived the need for patient consent as these archival tissues were no longer necessary for patient diagnostics. Researchers were not provided with any information pertaining to the identity of the patients.

***Data Availability Statement:*** *A subset of representative whole slide PARS, and chemical H&E image pairs have been uploaded to the BioImage Archive. This data is publicly available under accession number S-BIAD2324, accessible at https://doi.org/10.6019/S-BIAD2324. Data is provided in OME-TIFF format and* *can be viewed with standard pathology software such as QuPath. Other datasets used and analyzed during this study are available from the corresponding author on reasonable request.*

***Acknowledgments:*** The authors thank Drs. David Beyer, Karen Naert, Ami Wang, Pavandeep Gill, Xiao Zhu, and Calvin Tsengfor participating in the survey and questionnaire.

***Conflicts of Interest:*** The authors JEDT, BRE, HG, DD, JRM, and PHR all have financial interests in IllumiSonics, which has provided funding to the PhotoMedicine Labs. The authors MAD, AKS, MPW and GB do not have any competing interests. The assessment data were generated by pathologists masked to the image origin, and these pathologists have no financial interest in the outcomes of the study. The image order was randomized with guidance from MPW, the statistician analyzed the study; MPW also has no financial interests in the outcomes. All the data gathered were included in the analysis. The raw dataset, generated using SurveyMonkey, was directly provided to the statistician.

***Author Contribution:*** All authors performed review, read and approved the final manuscript. B.R.E and J.T performed study design, data collection, data processing. B.R.E performed virtual staining, writing, and revision. G.B., M.A.D., A.K.S., procured samples, performed writing and revision. G.B., and D.D., assisted in study design and sample selection. J.R.M., organized pathologist feedback and facilitated data collection. H.G., performed histological staining and sample processing. M.P.W., performed statistical analysis and organized study procedures. P.H.R., was principal investigator.

***Funding Sources:*** This research was funded by: Natural Sciences and Engineering Research Council of Canada (DGECR-2019-00143, RGPIN2019-06134, DH-2023-00371); Canada Foundation for Innovation (JELF #38000); University of Waterloo Startup funds; Centre for Bioengineering and Biotechnology (CBB Seed fund); illumiSonics Inc (SRA #083181); New frontiers in research fund – exploration (NFRFE-2019-01012); The Canadian Institutes of Health Research (CIHR PJT 185984).